\RequirePackage{fix-cm}
\documentclass[twocolumn,epjc3]{svjour3}  
\smartqed  
\RequirePackage[dvipdfmx]{graphicx}
\RequirePackage{latexsym}
\RequirePackage[numbers,sort&compress]{natbib}
\RequirePackage[colorlinks,citecolor=blue,urlcolor=blue,linkcolor=blue]{hyperref}
\usepackage{array}
\usepackage{braket}
\usepackage{bm}
\usepackage{subfigure}
\usepackage{verbatim}
\usepackage{wrapfig}
\usepackage{ascmac}
\usepackage{makeidx}
\usepackage{ulem}
\usepackage[version=3]{mhchem}
\usepackage{here}
\usepackage{comment}
\usepackage{tikz}
\usetikzlibrary{intersections, calc, arrows, patterns}
\usepackage{seqsplit}
\usepackage{mathcomp}
\journalname{Preprint Typeset}
\begin{document}

\title{
A simulation model investigation of neutron-oxygen inelastic scattering and subsequent nucleus deexcitation based on experimental data
}


\author{
Y.~Hino\thanksref{addr1, e1},
Y.~Ashida\thanksref{addr2},
T.~Tano\thanksref{addr3},
and
Y.~Koshio\thanksref{addr3}
}

\thankstext{e1}{e-mail: hino@post.kek.jp}


\institute{%
High Energy Accelerator Research Organization (KEK), Tsukuba, Ibaraki 305-0801, Japan \label{addr1} \and
Department of Physics, Tohoku University, Sendai, Miyagi 980-8578, Japan \label{addr2} \and
Department of Physics, Okayama University, Okayama, Okayama 700-8530, Japan \label{addr3}
}

\date{Received: date / Accepted: date}

\maketitle

\begin{abstract}
The nuclear interaction model plays an essential role in understanding neutrino-nucleus interactions in large-scale neutrino detectors.
For example, in the Super-Kamiokande experiment, systematic uncertainties regarding atmospheric neutrino interactions on oxygen limit the sensitivity to some physics studies, such as the diffuse supernova neutrino background search.
Reduction of such uncertainties necessitates an accurate modeling of nuclear reactions and subsequent nuclear deexcitation.
For this purpose, a detailed study was performed by comparing various combinations of simulation models with the newly released neutron experimental data.
From this study, it is found that the combination of INCL++ and NucDeEx implemented in a Geant4-based simulation shows a better agreement with the experimental data.
The presented result will be referred to improve the future physics studies at neutrino detectors, including the current focus of Super-Kamiokande. 
\end{abstract}

\section{Introduction}
\label{intro}
A comprehensive understanding of final-state particles resulting from neutrino-nucleus interactions is imperative for a variety of physics studies at large-scale neutrino detectors, such as nucleon decay search, CP-phase and mass ordering measurements in neutrino oscillations, as well as diffuse supernova neutrino background (DSNB) search.
In these measurements, GeV-scale neutrinos, e.g., atmospheric neutrinos, produce signal or background events via charged current (CC) or neutral current (NC) interactions on detector material.
Such neutrino interactions are likely to involve hadronic particles spilled out from nuclei.
Outgoing hadrons subsequently interact with nuclei in a detector, resulting in additional particle emission (nucleons and gamma-rays), which can affect the event topology and hence neutrino event reconstruction in physics analysis.

Super-Kamiokande (SK) is a large water Cherenkov detector consisting of 50 ktons of ultra-pure water, aiming at various physics targets.
Recently, SK has been upgraded by loading gadolinium (Gd) into pure water and started a new phase of experiment known as SK-Gd~\cite{beacom, first, second}.
In addition to representing the most significant cross section of thermal neutron capture among available elements, the total energy of gamma-rays generated by neutron capture on Gd is approximately 8 MeV. 
With the Gd-loaded water, a combination of a tight timing correlation and a high light yield of capture events results in an efficient improvement in the signal-to-noise ratio compared to the pure water phase.
For example, making a coincidence of the primary neutrino interaction and the delayed neutron signal achieves a significant accidental background rejection in the DSNB search~\cite{dsnb_sk4} and makes it possible to identify the wrong-sign component in neutrino oscillation measurement~\cite{wester}.
It can also work to reduce the atmospheric neutrino background in proton decay search, as no associated neutron is expected in many scenarios~\cite{takenaka}.

Despite the huge benefits of the neutron tagging, there is a concern about neutron-associated events in the DSNB search, particularly NC quasi-elastic (NCQE) interactions induced by atmospheric neutrinos. 
NCQE interactions on oxygen knock out a nucleon and leave an excited nucleus, which subsequently generates gamma-rays through deexcitation~\cite{ankowski}, as:
\begin{equation}
    \begin{split}
        \nu + {}^{16}\mathrm{O}& \to \nu' + {}^{15}\mathrm{O}^{*} + p, \\
        \nu + {}^{16}\mathrm{O}& \to \nu' + {}^{15}\mathrm{N}^{*} + n.
    \end{split}
\end{equation}
This channel was previously measured by the T2K neutrino beam without neutron tagging, and it was therein reported that the reconstructed Cherenkov angle distribution shows a discrepancy at large angles between the observed data and the expectation by Monte-Carlo (MC) simulation~\cite{ncqe_t2k}.
More recently, T2K performed a new NCQE study with neutron tagging on its initial dataset from the SK-Gd phase~\cite{ncqe_t2k_sk6}. 
This analysis provides intense comparisons between the observed data and various MC model predictions about the number of detected neutrons after NCQE. 
Here, a large discrepancy is reported, particularly for the model choice used in the previous T2K measurement~\cite{ncqe_t2k}.
Such discrepancies are thought to be attributable to an insufficient modeling of neutron inelastic scattering and the gamma-ray production process. 
In the DSNB search analysis at SK~\cite{dsnb_sk4, dsnb_sk6}, a large systematic uncertainty (68--82\% depending on the energy region) is assigned on the expected NCQE background due to the currently observed discrepancy as mentioned above. 
This, in turn, limits the search sensitivity. 
The situation should be improved by replacing the model with one that reproduces the actual neutron behavior in water.
The previous studies with SK's atmospheric neutrino samples~\cite{ncqe_sk6, atmnu_sk6} suggest a recovery of data-MC consistency by modifications to the nuclear scattering model. 

As indicated in the previous studies of neutrino interactions, an improvement to the nucleon-nucleus scattering model is a key to the precise prediction of the neutrino event topology and reconstruction at detectors.
In order to validate various nuclear scattering models, a simulation study was performed in this paper. 
Here, the newly released neutron-oxygen experimental data from the E525 experiment, which utilized 30 and 250~MeV quasi-monoenergetic neutron beams at Osaka University's Research Center for Nuclear Physics (RCNP), are referred~\cite{E525}. 
In the following part of the paper, comparison results among a variety of combinations of the intranuclear cascade model and the deexcitation model are presented. 
This will allow us to select a model that provides a superior prediction based on the experimental data, as well as to improve the model for further precision.

It should be noted that, ahead of E525, the E487 experiment was performed on a similar concept using an 80 MeV neutron beam at the same facility~\cite{E487}. While E487 has established the experimental methodology and successfully observed multiple gamma-ray signals induced by neutron-oxygen reactions, there might be an issue with the injected proton current measurement, which changes the overall scale of the measured production cross section, as pointed out in detail in Ref.~\cite{E525}. 
We therefore decided to focus on the E525 results for comparison with our simulation in this paper. Nevertheless, because the E487 result is still useful and can be recovered, we leave this in the appendix of the paper. Such recovered results will be compared with our simulated sample and discussed therein.

\section{Experimental Dataset}
\label{sec:dataset}
The objective of the E525 experiment was to ascertain the uncertainty derived from the neutron-nucleus reaction associated with the NCQE interaction.
To this end, the experiment measured gamma-rays resulting from the neutron-oxygen reaction at neutron kinetic energies of 30 and 250 MeV in two separate beam tests.
Both tests were conducted at RCNP, utilizing the neutron time-of-flight beamline.
A water target, which is composed of a cylindrical acrylic vessel filled with water, was irradiated with a quasi-monoenergetic neutron beam produced by the ${}^7$Li($p$, $n$)${}^7$Be reaction.
It is noteworthy that two water vessels were utilized in the test with the 250 MeV neutron beam, in contrast to the single vessel employed in the 30 MeV neutron beam test.
The gamma-ray spectra were measured using a high-purity germanium (HPGe) semiconductor detector.
The configurations of each test are enumerated in Table~\ref{tab:exp_summary}.
The gamma-ray production cross sections at each beam neutron energy were measured by fitting the observed spectra.
Concurrently, such spectrum data serve as an essential validation benchmark for simulation models of neutron inelastic interaction~\cite{E525}.

\begin{table}[htbp]
    \centering
    \caption{Summary of the experimental setup configuration in each test of the E525 experiment. ``Energy'' shows the parent proton beam energy and ``Volume'' corresponds to the target water volume.}
    \begin{tabular}{ccc}
        \hline
        Test\# & Energy /MeV & Volume /L   \\ \hline
        1       & 30          & 6.9                 \\
        2       & 250         & 13.9                \\
        \hline
    \end{tabular}
    \label{tab:exp_summary}
\end{table}

\section{Simulation}
\label{simulation}

\subsection{Setup}
\label{sec:setup}

To validate simulation models, a dedicated MC simulation was tailored with the experimental setup of the E525 experiment based on the Geant4 simulation toolkit ~\cite{geant4_1,geant4_2,geant4_3}.
The utilization of Geant4.10.5.p01 was specifically driven by its compatibility with the version that the latest SKG4, Geant4-based SK detector simulation~\cite{skg4}, depends on.

\begin{figure}[htbp]
    \centering
    \includegraphics[width=0.95\linewidth]{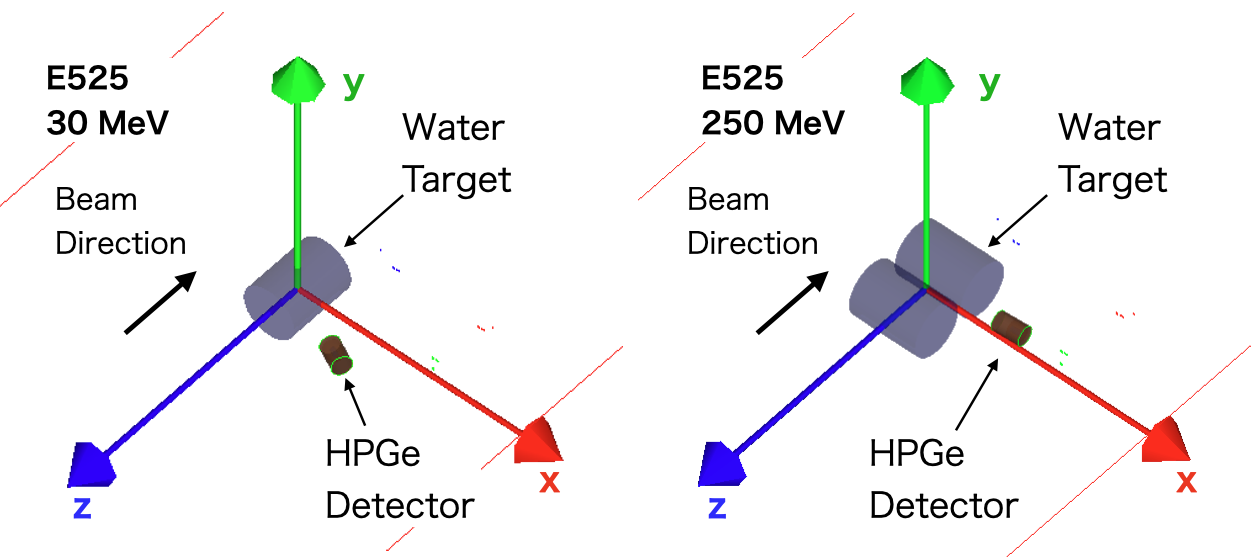}
    \caption{The visualized simulation setups of the test using the 30~MeV beam (left) and the 250~MeV beam (right).}
    \label{fig:sim_setup}
\end{figure}

Figure~\ref{fig:sim_setup} illustrates the implemented experimental setup in the MC simulation.
It is important to note that the position of the gamma-ray detector was approximately 1 m away from the water target.
As mentioned in Section~\ref{sec:dataset}, the two identical water vessels were used in the test with a 250 MeV neutron beam.
The neutron beam was generated based on the measured beam profile in each experiment.
Given that the analyses of the experimental data were performed by selecting the peak corresponding to the monoenergetic neutron, the target was simply irradiated with neutrons having kinetic energies of 30 and 250 MeV in the present MC simulations.

Energy deposition in the HPGe detector was converted to visible energy by the energy resolution response, which was estimated from the calibration data.
A validation of the gamma-ray detector in the MC simulation based on the ${}^{60}$Co source demonstrated that the observed energy spectra were consistent within 5\% with the corresponding simulation.

\subsection{Physics model}
\label{sec:inelamodel}

In this study, we tested multiple models compatible with Geant4. 
In the energy region of current interest, e.g., 20 MeV to 1 GeV for NCQE-induced neutrons in SK, the final state products of a single hadron inelastic interaction are determined sequentially by a combination of intranuclear cascade and deexcitation models.
As a cascade model, Bertini cascade (BERT), Binary cascade (BIC), and the Liege intranuclear cascade (INCL++) are implemented and available in Geant4.
Note that the model causing the large uncertainty in the NCQE prediction in SK is based on GCALOR in Geant3~\cite{gcalor}, which is essentially identical to BERT in Geant4. 

The cascade model is generally used to determine products of interactions through a cascade of binary collisions between a projectile and nucleons within a target nucleus~\cite{bertini}.
Pauli blocking is an inherent property of binary particle-particle collisions.
The emission of hadrons and light clusters follows a theoretical possibility. 
In particular, light clusters are produced via a dynamical phase-space coalescence algorithm.
The cascade is terminated when all cascade particles have either departed or become trapped in a nucleus, as observed in the BERT and BIC models.
On the other hand, INCL++ terminates the cascade by the time self-consistently defined as $t_{\mathrm{stop}} = t_0 \left( A_{\mathrm{target}}/{208} \right)^{0.16}$,  where $t_0$ is a parameter set to 70 fm/c and $A_{\mathrm{target}}$ is the mass number of the target nucleus~\cite{incl}. 

After the cascade stage, a pre-equilibrium phase of nuclear reaction ensues, persisting until the system attains equilibrium by undergoing state transitions accompanied by light particle emission up to $\alpha$.
This treatment is implemented in a part of BERT based on the statistical Griffin exciton model~\cite{griffin}.
For the semi-classical exciton model based on the Griffin model, the implementation of both BIC and INCL++ occurs within \seqsplit{G4PreCompoundModel} as a backend.
Consequently, there is negligible variation among the three cascade models at this particular point.

At the equilibrium stage, the residual nucleus in the excited state relaxes via particle emission (mostly gamma-rays and nucleons, in the case of oxygen), fission, or other mechanisms.
To simulate the deexcitation process, the cascade models are coupled with one of the separate statistical deexcitation codes.
The native evaporation model implemented in BERT is based on the Weisskopf statistical theory~\cite{weisskopf}, which computes the emission of particles until the excitation energy reaches a cutoff~\cite{g4bert}. 
In the process of photon evaporation, for example, successive emissions of gamma-rays are observed, with the energy of each emission derived from an exponential-like function. This process continues until the excitation energy reaches a threshold value, defined by $E^{*} < E^{\gamma}_{\it cutoff} = 10^{-15}$ MeV~\cite{dostrovsky}.
In the implementation, no discrete gamma-ray energy corresponding to nuclear levels is considered.
It is important to note that the process of particle emission resulting from Fermi break-up does not apply to oxygen and carbon due to the limitation on the mass number, $A < 12$.

Conversely, the default deexcitation model of BIC and INCL++ consists primarily of G4Evaporation (for particle emission) and \seqsplit{G4PhotonEvaporation} (for gamma-ray emission), which are utilized in \seqsplit{G4PreCompoundModel}.
The primary function of the model is to compute particle emission up to $\alpha$ particle based on the Weisskopf-Ewing model in the case of oxygen and carbon~\cite{precompound}.
It should be noted that alternative candidates for the deexcitation model are available for INCL++.

ABLA++, a translation to C++ of the Fortran-based ABLA07, simulates the deexcitation by calculating the probabilities for emitting gamma-rays, neutrons, light-charged particles, and intermediate mass fragments according to Weisskopf's formalism~\cite{abla}.

Another candidate is NucDeEx~\cite{nucdeex, nucdeex_v2}, recently developed as a nuclear deexcitation event generator for neutrino interactions and nucleon decay.
It has been demonstrated to manage particle emission in scenarios where excitation energy exceeds separation energy and multi-hole states, contingent on the branching ratios calculated by the TALYS code~\cite{talys}.
TALYS employs the Hauser-Feshbach statistical model~\cite{hfmodel} to compute branching ratios considering angular momentum conservation, in contrast to other models based on the Weisskopf theory.
Gamma-ray emission is called if the excitation energy is between the first excited state and the separation energy.
The excited state and its branching ratio employed in NucDeEx are based on the experimental data, e.g., ${}^{15}\mathrm{N}^{*}$ and ${}^{11}\mathrm{B}^{*}$.
In contrast, the branching ratios for ${}^{15}\mathrm{O}^{*}$ and ${}^{11}\mathrm{C}^{*}$ are assumed to be the same as those for ${}^{15}\mathrm{N}^{*}$ and ${}^{11}\mathrm{B}^{*}$, respectively.
This assumption is made in the context of isospin symmetry, except for the Coulomb potential difference between protons and neutrons, due to the absence of experimental data in this area.

\begin{figure}[htbp]
    \centering
    \includegraphics[width=0.95\linewidth]{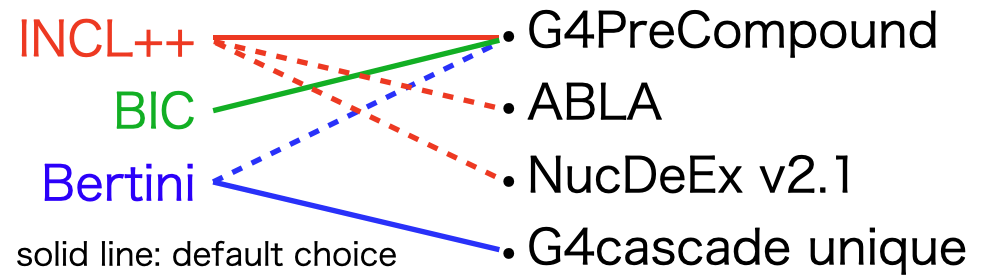}
    \vspace{+10truept}
	\caption{Possible combinations of intranuclear cascade and deexcitation models compatible with Geant4. The solid lines indicate the default choice of each physics list, and the dashed lines display other tested choices.}
    \label{fig:models}
\end{figure}

A summary of the combinations of intranuclear cascade and deexcitation models which we investigated is shown in Fig.~\ref{fig:models}.
Note that neutron reactions below 20~MeV are managed by G4ParticleHP, enabling the high precision neutron transportation based on the evaluated nuclear database.
G4NDL4.5, which primarily refers to ENDF/B-VII.1, was utilized in this study.

\section{Result}
In this section, a comparative analysis is conducted between simulation outputs from different pairs of intranuclear cascade and deexcitation models and the experimental dataset provided from the E525 experiment.
The reproducibility of the MC simulation results is assessed through the implementation of a chi-squared test on the observed gamma-ray spectra within the HPGe detector.
The neutron flux uncertainty of each experiment is considered a pull term in the chi-squared evaluation as:
\begin{equation}
    \chi^2 = \sum^{bins}_{i} \left(\frac{X^{\mathrm{Data}}_{i}-\eta \times X^{\mathrm{MC}}_{i}}{\sigma^{\mathrm{stat.}}_{i}}\right)^2 + \left( \frac{1-\eta}{\sigma^{\mathrm{flux}}} \right)^2,
    \label{eq:chi2}
\end{equation}
where $X_{i}$ is the count at $i$-th bin of the spectra, $\sigma^{\mathrm{stat.}}$ is the statistical uncertainty of the data, $\eta$ is the nuisance parameter representing the pull term constrainned by the neutron flux uncertainty $\sigma^{\mathrm{flux}}$ of each experiment.

\subsection{Standard Geant4 models}
\label{cascade}
\begin{figure*}[tb]
    \centering
    \includegraphics[width=0.46\linewidth]{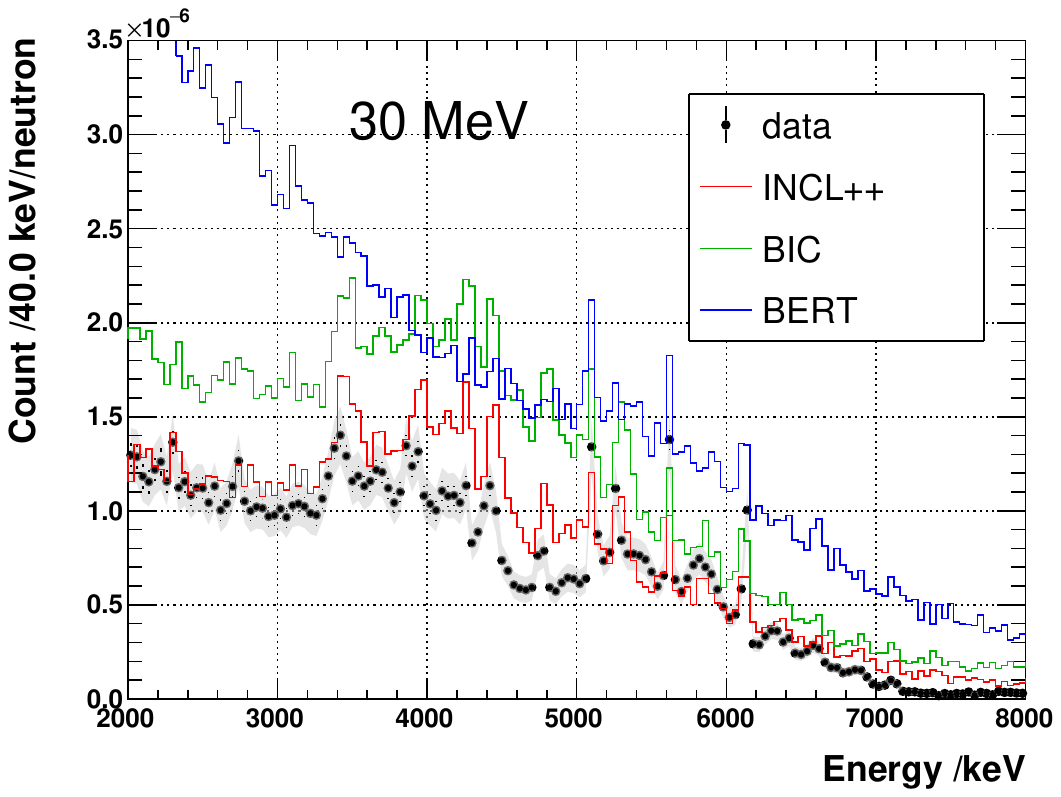}
    \includegraphics[width=0.46\linewidth]{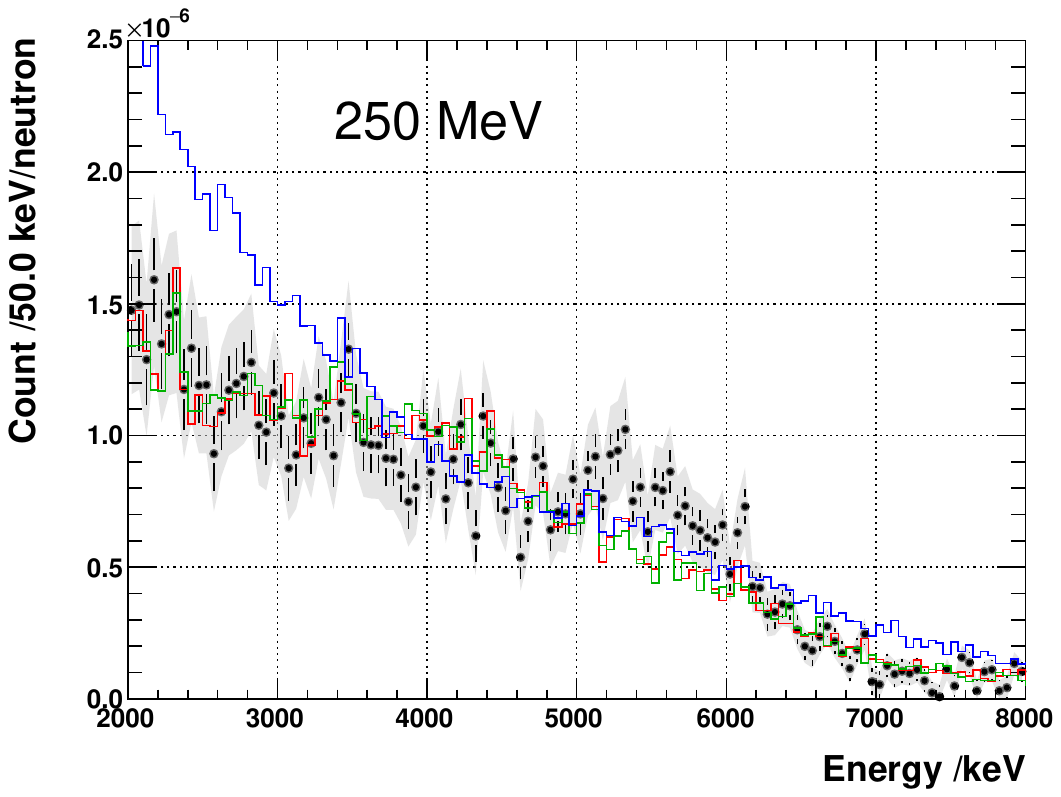}
    \includegraphics[width=0.46\linewidth]{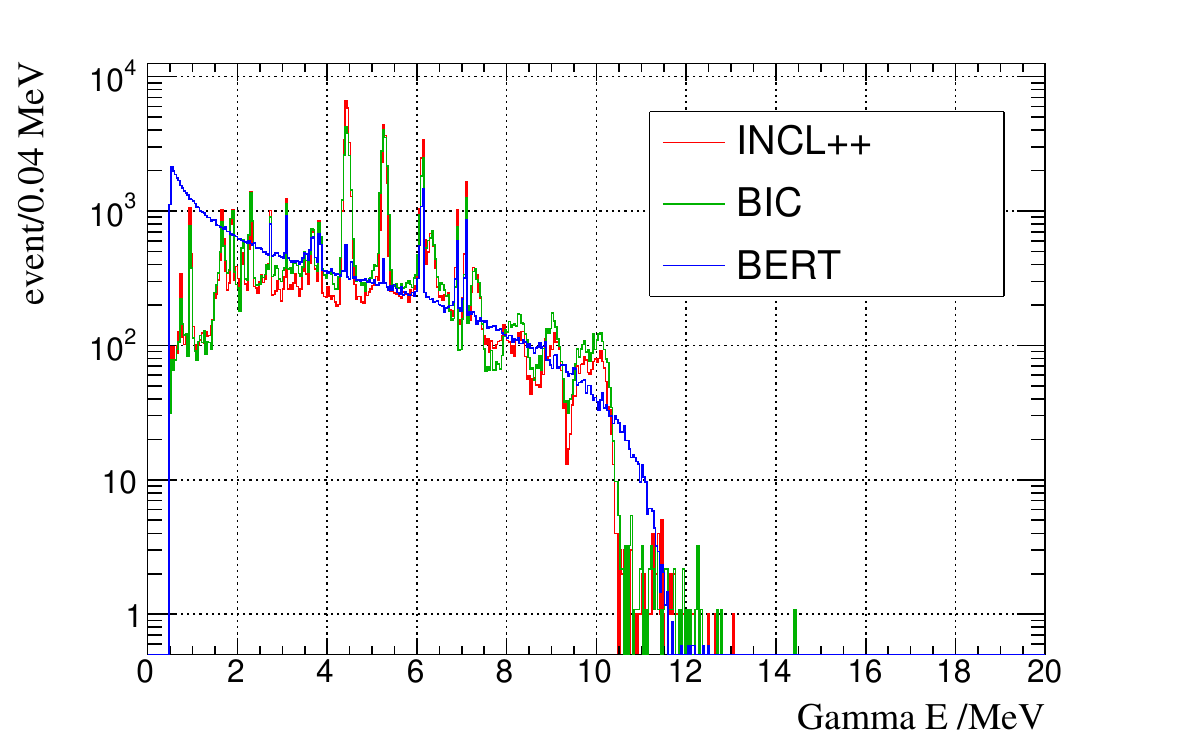}
    \includegraphics[width=0.46\linewidth]{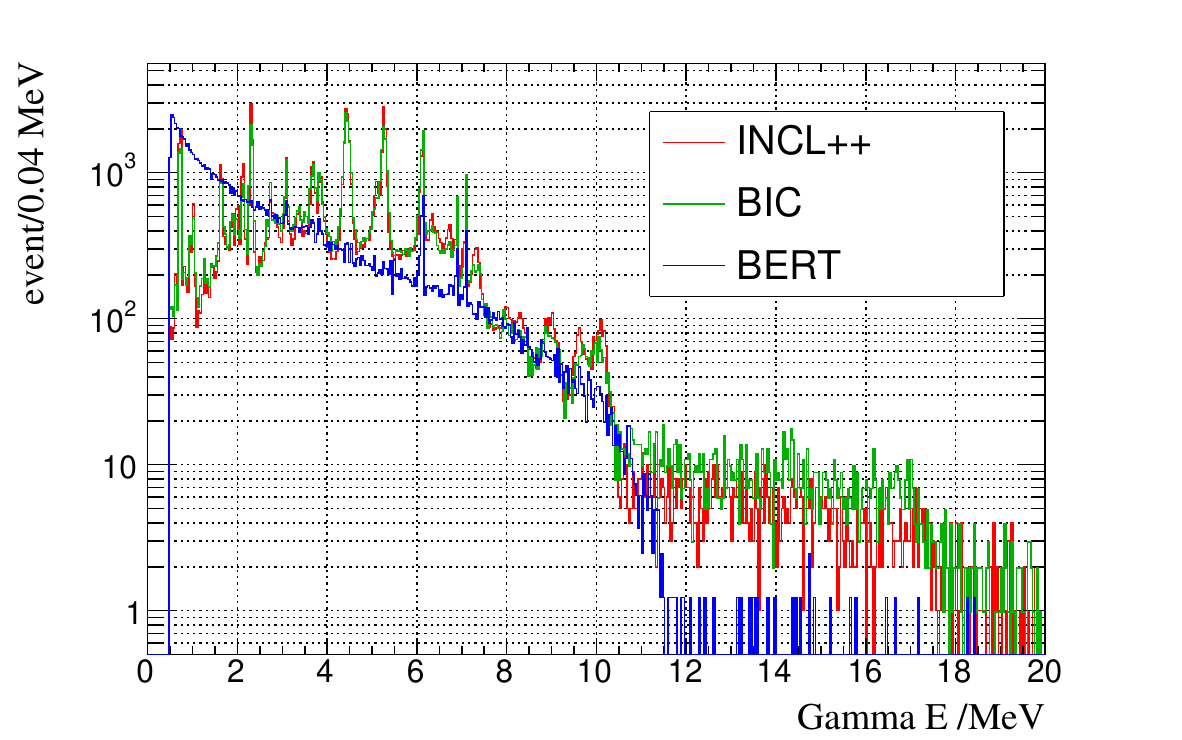}
    \vspace{+5truept}
	\caption{Top panels: The observed energy spectra of gamma-rays at 30 MeV (left) and 250 MeV (right) of the beam energy. Each plot shows the E525 data (black marker) overlaid with the MC simulations, including BERT (blue), BIC (green), INCL++ (red), with their default deexcitation model choice. The gray band, along with the data points, corresponds to the systematic uncertainty in the measurement. Bottom panels: The true generated energy spectra of gamma-rays from neutron-oxygen interaction at 30 MeV (left) and 250 MeV (right) of the beam energy. }
    \label{fig:cascade}
\end{figure*}

Initially, a comparison is made between the Geant4 default pairs of the cascade and deexcitation model and the experimental data.
In the top two panels of Fig.~\ref{fig:cascade}, the gamma-ray energy spectra of the data (black) and the MC simulations using the models (red: INCL++, green: BIC, blue: BERT) are displayed for two different beam energies, 30 MeV (left) and 250 MeV (right).
It also shows the energy distributions of the generated gamma-rays at each neutron beam energy in the bottom two panels.

Table~\ref{tab:cascade} shows a summary of the chi-squared evaluation in each cascade model.
According to the definition of the chi-squared in Eq.~\ref{eq:chi2}, the chi-squared value is indicative of reproducibility in the spectrum shape, and the magnitude of the pull signifies the precision of the absolute gamma-ray production by the neutron-nucleus interaction.
For the beam energy of 30~MeV, the performance of INCL++ and BIC in terms of spectral shape is comparable, despite the disparity in the magnitude of the pull term.
BERT demonstrated a tendency to overestimate the gamma-ray production with a continuum component in an exponential function over the entire energy range.
This overestimation led to an increase in both the chi-squared statistic and the pull.
In the instance of the 250 MeV beam energy, these tendencies were compromised.
This can also be found in the energy spectrum of the true generated gamma-rays at 250 MeV, which shows less significant peak intensities in INCL++ and BIC than at 30 MeV.
Both INCL++ and BIC predicted almost identical spectral shape and production of gamma-rays, and demonstrated good agreement with the data as shown in the right panel of Fig.~\ref{fig:cascade}.
Although there were inconsistencies in the spectral shape due to the continuum contribution in BERT, the dependence of the production rate on the beam energy converged among the models.
The behavior of BERT is natural by considering the implementation of gamma-ray deexcitation as described in Section~\ref{sec:inelamodel}.

In summary, the INCL++ model demonstrates superior performance in predicting gamma-ray production from neutron-oxygen interactions.
As previously indicated, there are certain customizable options for the deexcitation model that are compatible with INCL++ in Geant4.
This topic will be examined in the subsequent section.

\begin{table}[bt]
    \centering
    \caption{The chi-squared and pull term values of each model to the data from the E525 experiment. The degrees of freedom are 150 in the 30~MeV case and 120 in the 250~MeV case, respectively.}
    \begin{tabular}{ccc} \hline
        Model         & $\chi^2_{30}/dof$ (pull) & $\chi^2_{250}/dof$ (pull) \\ \hline
        INCL++ &  6.9 (2.9) & 2.57 (0.05) \\
        BIC    &  6.9 (16.8) & 2.91 (0.05) \\
        BERT   & 10.0 (34.8) & 4.94 (0.89) \\ \hline
    \end{tabular}
    \label{tab:cascade}
\end{table}

\subsection{Alternative deexcitation models}
\label{deexcitation}
The comparison between the cascade models coupled with their default deexcitation model indicates that the differences in the deexcitation model affect the spectral shape.
Here, the other options of the deexcitation model compatible with INCL++ are tested.
Note that a NucDeEx-based deexcitation module compatible with INCL++ in Geant4 is specially provided by the NucDeEx developer in private communication.

\begin{figure}[bt]
    \centering
    \includegraphics[width=0.99\linewidth]{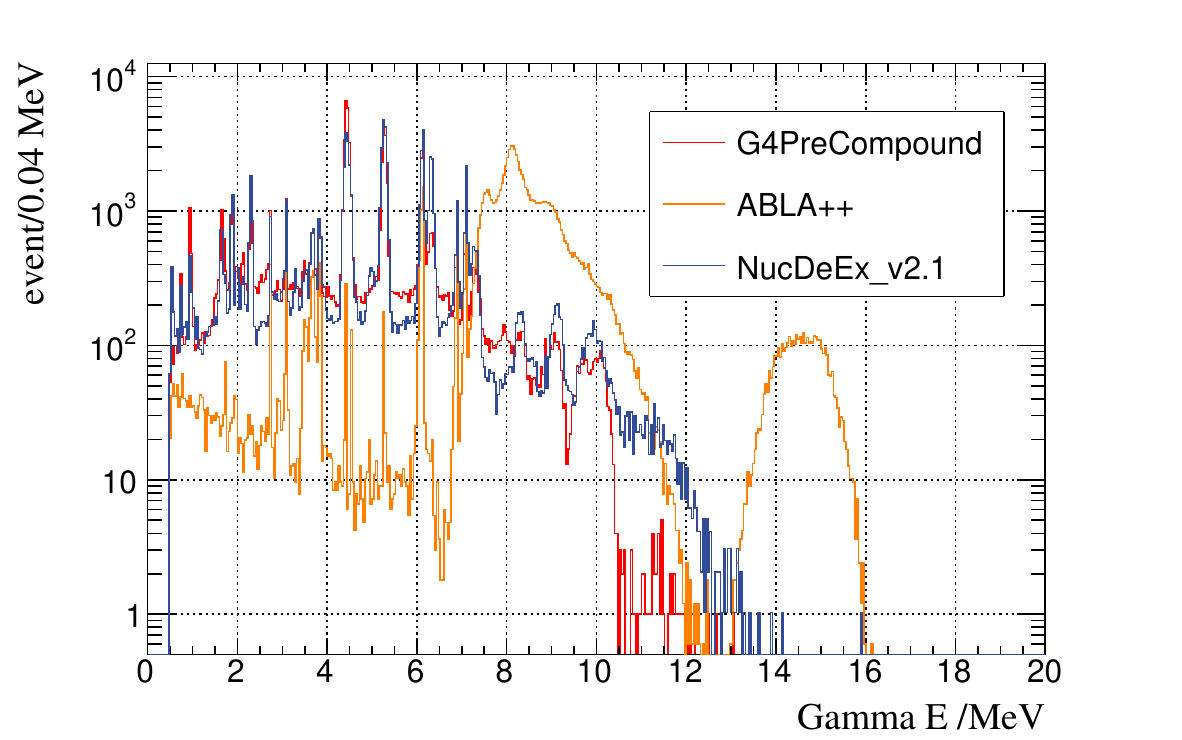}
	\caption{The true generated gamma-ray energy spectra of the MC simulation with G4PreCompoundModel (red), NucDeEx (dark blue), and ABLA++ (orange) coupled with INCL++ at 30 MeV of the beam energy. Only ABLA++ showed a largely different spectrum to the others, while the combination of INCL++ and G4PreCompoundModel demonstrated a good agreement with the E525 data in Fig.~\ref{fig:cascade}.}
    \label{fig:deex_gen}
\end{figure}

\begin{figure*}[bt]
    \centering
    \includegraphics[width=0.46\linewidth]{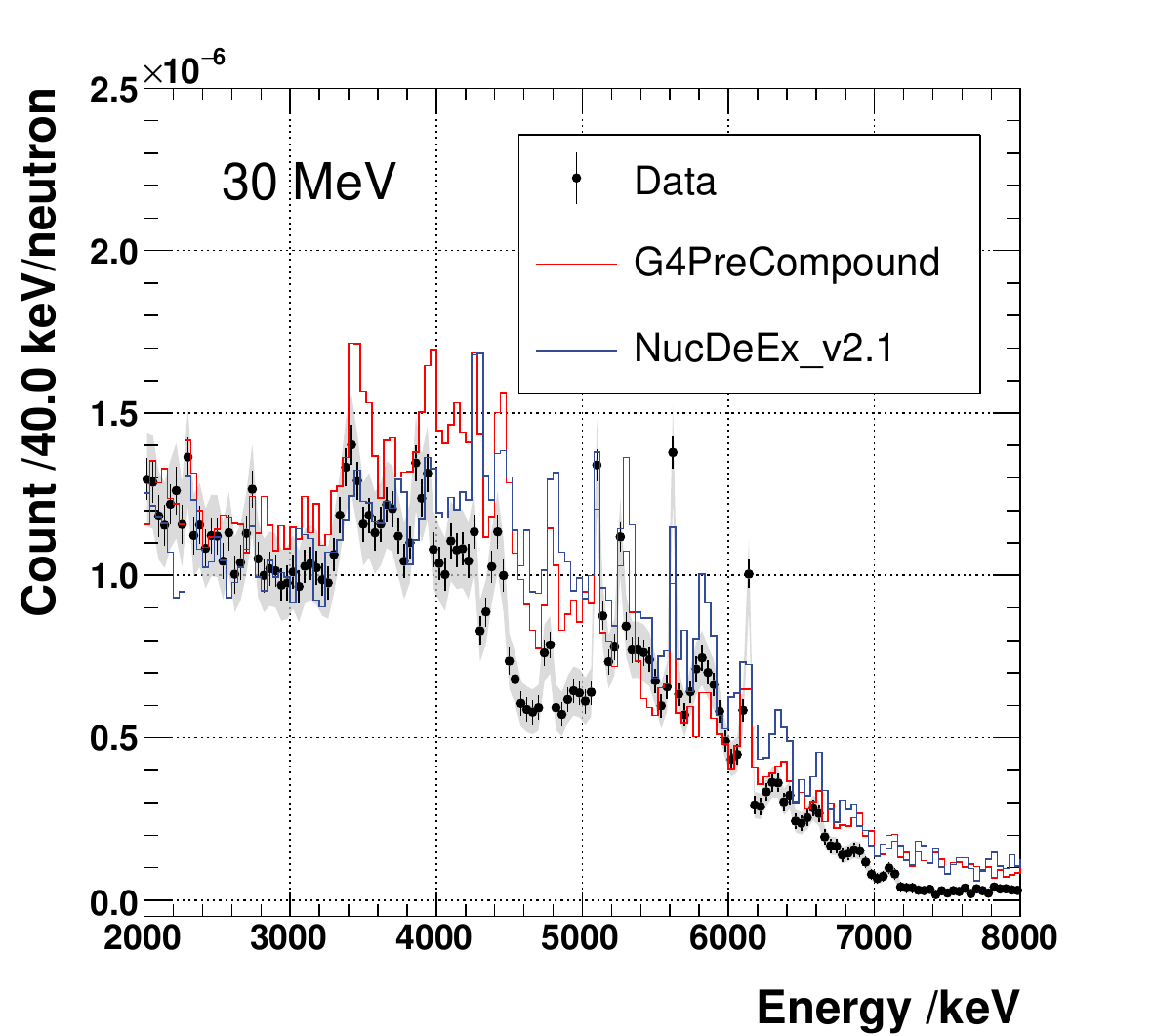}
    \includegraphics[width=0.46\linewidth]{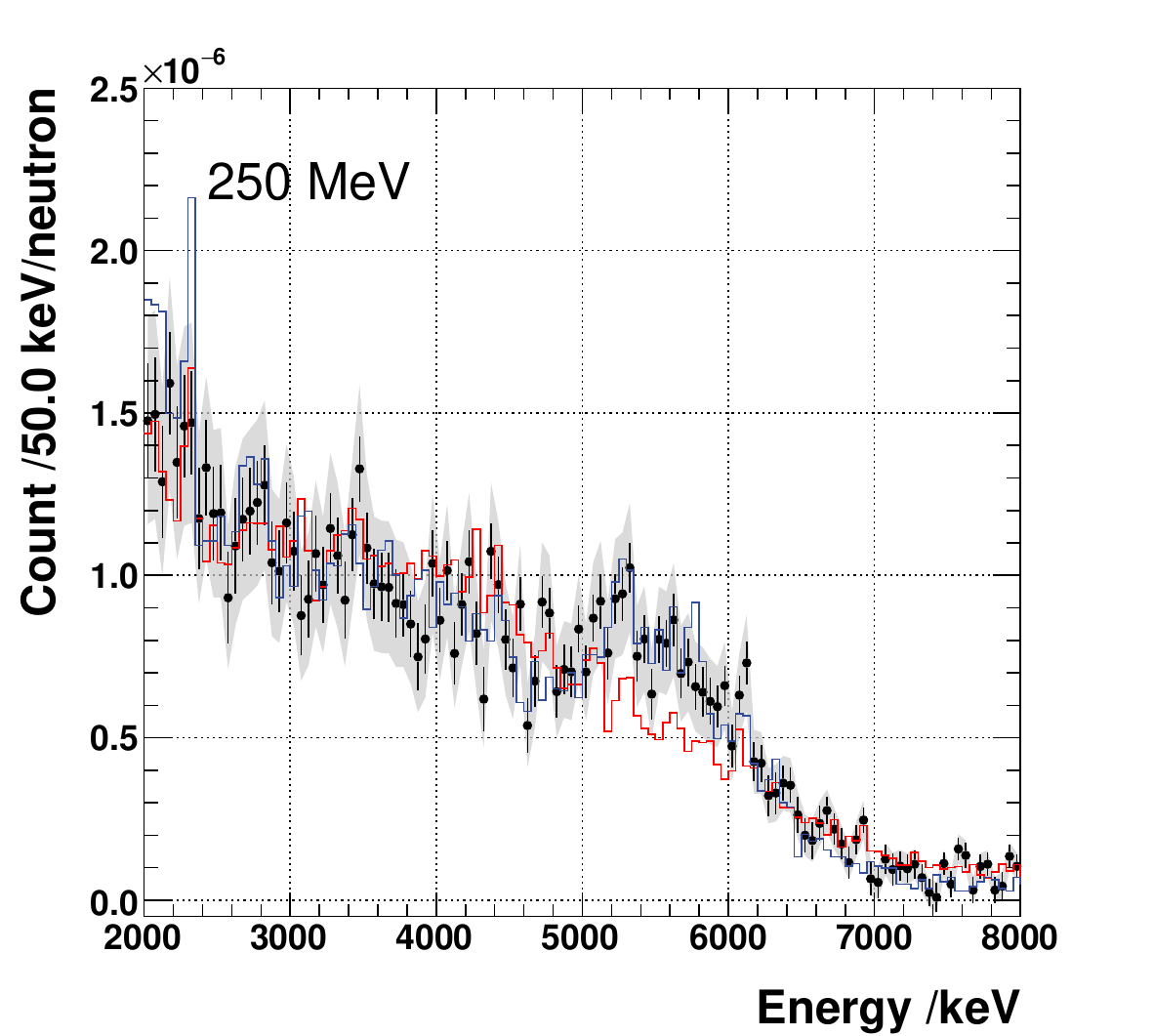}
    \vspace{+5truept}
	\caption{The observed energy spectra of gamma-rays from neutron-oxygen interaction at 30 MeV (left) and 250 MeV (right) of the beam energy. Each plot compares that of the E525 experimental data (black marker) and the MC simulations with G4PreCompoundModel (red) and NucDeEx ver2.1 (dark blue) coupled with INCL++. The gray band, along with the data points, corresponds to the systematic uncertainty in the measurement. Although both show a similar performance in terms of the gamma-ray production, NucDeEx shows better agreement at both beam energies.}
    \label{fig:deex}
\end{figure*}

Primarily, we compared energy spectra of true generated gamma-rays in neutron-oxygen interaction from different deexcitation models as shown in Fig.~\ref{fig:deex_gen}.
A comparison of the spectra predicted by \seqsplit{G4PreCompoundModel} (red) and NucDeEx (dark blue) reveals a high degree of similarity, except for minor discrepancies.
However, a significant disparity is apparent when ABLA++ is considered, exhibiting substantial variations across the entire spectrum.
The present study has demonstrated that the ABLA++ model overpredicts the contribution of the gamma-ray energy at $>$8~MeV. This finding is inconsistent with the observed consistency between G4PreCompoundModel and the experimental data shown in Fig.~\ref{fig:cascade}.
In the following discussion, ABLA++ is not considered in our investigation.

Finally, the experimental data at the beam energies of 30 and 250 MeV were compared with the MC simulations with different deexcitation models.
Figure~\ref{fig:deex} shows the gamma-ray energy spectrum of the data (black) and the MC simulations using \seqsplit{G4PreCompoundModel} (red) and NucDeEx (dark blue) coupled with INCL++ at 30~MeV (left) and 250~MeV beam energy (right), respectively.
\begin{table}[bt]
    \centering
    \caption{The chi-squared values of each model for the data from the E525 experiment. The degrees of freedom are 150 (30~MeV) and 120 (250~MeV), respectively.}
    \begin{tabular}{ccc} \hline
        Model         & $\chi^2_{30}/dof$ (pull) & $\chi^2_{250}/dof$ (pull) \\ \hline
        G4PreCompound &  6.9 (2.9) & 2.57 (0.05) \\
        NucDeEx v2.1  &  6.2 (2.9) & 1.67 (0.01) \\ \hline
    \end{tabular}
    \label{tab:deex}
\end{table}
Table~\ref{tab:deex} shows a summary of the chi-squared evaluation based on Eq.~\ref{eq:chi2} in each model.
Because the pull values were similar between the results due to the common cascade model (INCL++), the chi-squared values were only displayed in order to compare the spectral shapes mainly.
NucDeEx shows better agreement with the data at each beam energy than \seqsplit{G4PreCompoundModel}.
In particular, clear differences can be seen between the models in the range of 5,000 and 6,000~keV, resulting in the large chi-squared value.
The gamma-ray from the 3/2-hole state of ${}^{15}\mathrm{N}$ (6.32 MeV) contributes to this range by its single and double escape components~\cite{E487, E525}.
This validates the performance of NucDeEx as a deexcitation simulator worth using as a replacement for the models available in Geant4.

\section{Conclusion and Discussion}
\label{conclusion}
A simulation study of neutron-oxygen interaction was performed to improve the precision of the neutrino interaction prediction.
We compared the simulations in the various settings with the newly available measurements of gamma-ray production in neutron-oxygen reactions from the E525 experiment.

As a result, BERT, the long-standing hadron inelastic model, was found to predict a larger amount of gamma-ray production than that of the experimental data.
This reasonably describes a cause of the discrepancy in the Cherenkov angle distribution in the NCQE measurements~\cite{ncqe_t2k, ncqe_sk6} because the Cherenkov angle of events with multiple gamma-rays tends to be reconstructed as larger.
On the other hand, BIC and INCL++ showed better agreement with the data if they utilize \seqsplit{G4PreCompoundModel} as a deexcitation model and predicted the smaller gamma-ray production than BERT.
Therefore, the simulation model using \seqsplit{G4PreCompoundModel} can improve the prediction of the NCQE reaction in the way of event topology. 

In addition, further investigation was carried out, concentrating on the other deexcitation models compatible with INCL++ in Geant4.
ABLA++ deexcitation model gave a large difference in the gamma-ray spectrum from others, while \seqsplit{G4PreCompoundModel} showed reasonable consistency with the data.
Finally, we confirmed that a combination of NucDeEx and INCL++ demonstrated the best agreement with the experimental data in terms of the production rate and the spectral shape.
As discussed in Ref.~\cite{nucdeex}, NucDeEx shows the best or comparable prediction in the aspect of particle emission from deexcitation in addition to the gamma-ray emission spectra.
Therefore, it is anticipated that uncertainties on the neutrino interaction will be improved with the better prediction of the final state particles from nuclear deexcitation in both primary neutrino-nucleus and secondary nucleon-nucleus interactions by NucDeEx.

By focusing on gamma-ray production in neutron-oxygen scattering, we concluded that MC poorly reproduced the event topology of NCQE measurements with a certain model.
The substituting simulation with different intranuclear and deexcitation models, i.e., INCL++ and NucDeEx, shows better consistency in both the absolute production rate and the gamma-ray spectrum from neutron-oxygen interactions. 
This will address a precise prediction of NCQE background contamination in the DSNB signal region from the constraint in the large-angle region (78--$90\tcdegree$) of the reconstructed Cherenkov angle.
Thus, it is expected that a precise prediction of the NCQE events with the models we found leads to an improvement in the sensitivity of the search at large water Cherenkov detectors toward a discovery.

\begin{acknowledgements}
\sloppy
We are grateful to Seisho Abe of the University of Tokyo for our constructive discussion about the deexcitation model and the implementation of his deexcitation model, NucDeEx, in Geant4.
We also appreciate the E487 members having a discussion and helping to revisit their results for this work.
This work was supported by Japan MEXT KAKENHI Grant Number 26400292 and 20H00162.
\end{acknowledgements}

\appendix
\section{Discussion with the E487 Experiment}

Ref.~\cite{E525} pointed out that there was a potential issue with the injected proton beam current measurement in E487, which might result in a wrong normalization of neutron flux.
However, as mentioned in Sec.~\ref{intro}, it is still useful to compare the gamma-ray spectral shape between the observed data and the simulation at 80 MeV of the neutron energy, because this energy overlaps with the major range of the energy distribution of neutrons induced by the neutrino-oxygen NCQE interaction.

The previous measurement~\cite{iwamoto} reported that the neutron production cross section in the ${}^7$Li($p$, $n$)${}^7$Be reaction is stable within 10\% over the proton energy range of 80 to 389 MeV.
Therefore, we here decided to refer to the neutron flux at 250~MeV measured in E525 ($5.01 \times 10^{9}$ neutron/sr/$\mu$C) in our simulation, instead of the original flux reported by E487 ($1.87 \times 10^{10}$ neutron/sr/$\mu$C), and compared the resulting gamma-ray spectrum with the E487 data.
Note that this treatment is justified only because our current purpose is the shape comparison, not the absolute strength measurement.
Further investigations on the cause for this neutron flux overestimation from the E487 collaboration are appreciated.

\begin{figure}[bt]
    \centering
    \includegraphics[width=0.9\linewidth]{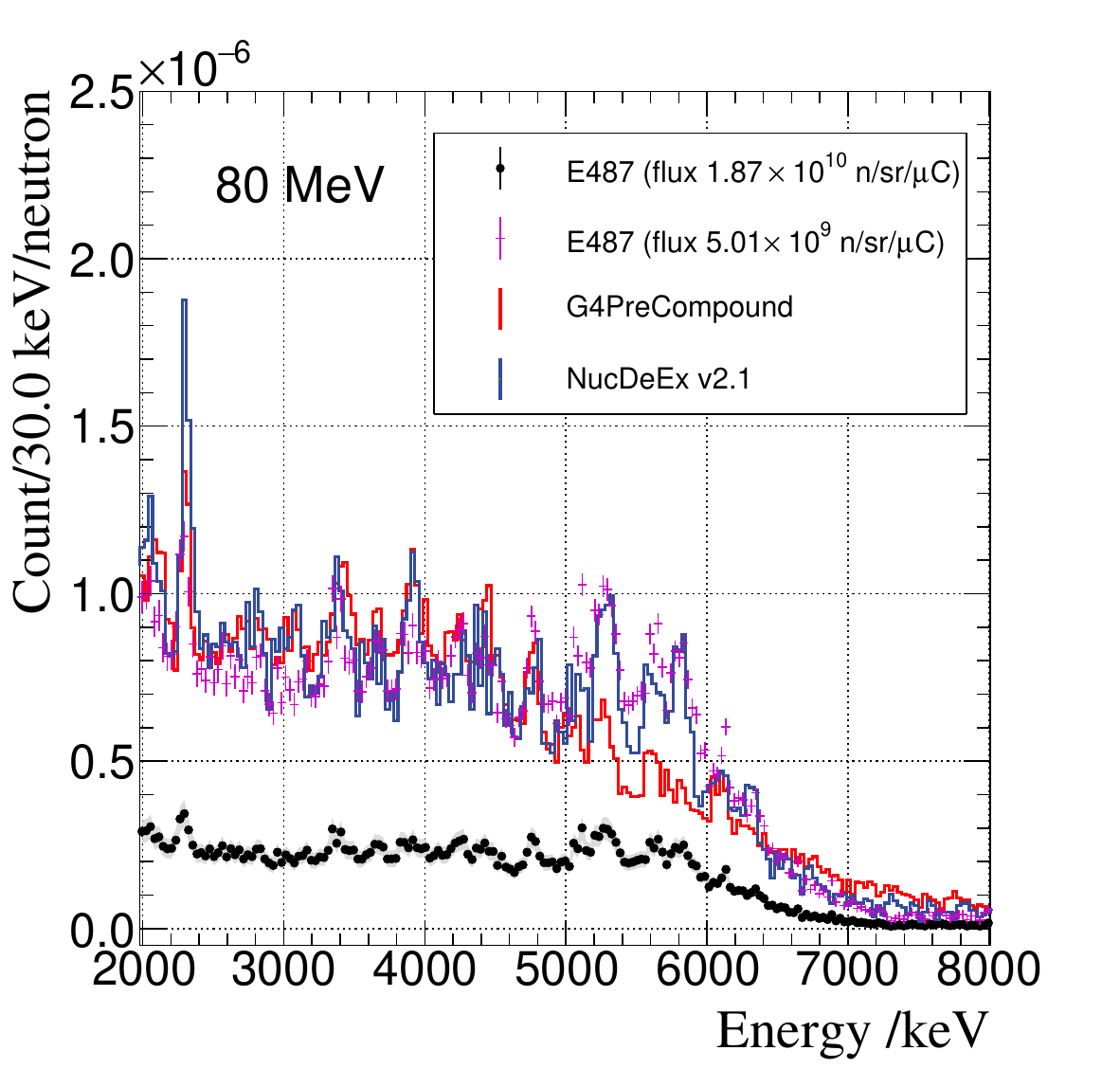}
    \vspace{+5truept}
	\caption{The observed gamma-ray energy spectra of the reported data by the E487 experiment (black markers) as well as the data points normalized with the E525 neutron flux at 250 MeV (magenta markers). The spectra predicted by NucDeEx (dark blue) showed better consistency in the spectral shape over 5 MeV than those predicted by G4PreCompoundModel (red).}
    \label{fig:flux_fit}
\end{figure}

In the E487 experiment, the neutron beam energy was 80 MeV, and the gamma-ray resulting from inelastic scattering on oxygen was observed using a lanthanum bromide scintillator, LaBr$_3$(Ce).
The experimental setup for E487 was similar to the E525 30~MeV setup, except for the gamma-ray detector.
We constructed the Geant4-based simulation, likewise the study shown in the main part.

Figure~\ref{fig:flux_fit} shows the E487 data with the alternative neutron flux from the E525 measurement (magenta) in comparison with the original E487 data (black) and the simulation results based on \seqsplit{G4PreCompoundModel} (red) and NucDeEx v2.1 (blue).
In the investigation of the simulation model, the NucDeEx prediction shows a better consistency with the E487 data than the \seqsplit{G4PreCompoundModel} one, as indicated by the smaller reduced chi-squared value (2.3 vs. 14.3).
Thus, in combination with the results presented in the main part of this paper, we concluded that NucDeEx coupled with INCL++ shows the best performance as the simulation model predicting neutron-oxygen interaction and deexcitation of residual nucleus at three different neutron energies covering the energy range of outgoing neutrons from the NCQE interaction.   



\end{document}